\pdfoutput=1
\documentclass[natbib]{aastex631} 

\usepackage{hyperref}
\defcitealias{CS15}{\scshape CS15}
\defcitealias{P22}{\scshape P22}
\defcitealias{PKT23}{\scshape PKT23}

\usepackage{amsmath} \usepackage[nointegrals]{wasysym}
\shorttitle{Toroidal Magnetic Flux Budget in Mean-Field Dynamo Model} \shortauthors{Pipin \& Kosovichev}

\begin{document}
	
	\title{Toroidal Magnetic Flux Budget in Mean-Field Dynamo Model of Solar Cycles 23
		and 24}
\author[0000-0001-9884-1147]{Valery V.~Pipin}
	\affiliation{Institute of Solar-Terrestrial Physics, Russian Academy of Sciences,
		Irkutsk, 664033, Russia}
	\author[0000-0003-0364-4883]{Alexander G. Kosovichev}
\affiliation{Department of Physics, New Jersey Institute of Technology, Newark, NJ 07102}
\affiliation{NASA Ames Research Center, Moffett Field, Mountain View, CA 94040}
	\correspondingauthor{Valery V. Pipin}
\email{valery.pipin@gmail.com}

	\begin{abstract}
 We study the toroidal magnetic flux budget of the axisymmetric part of a 3D mean-field dynamo model of Solar Cycles 23 and 24.
{The model simulates the global solar dynamo that includes effects of  the formation and evolution of bipolar magnetic
regions emerging on the solar surface. 
 Our  analysis shows that the hemispheric magnitude of  the net axisymmetric toroidal magnetic field in the bulk of the convection zone is partly defined by the surface parameters  of the differential rotation and  the axisymmetric radial magnetic field.  The contribution of the rotational radial shear to the net axisymmetric toroidal field production has the same magnitude and it goes nearly sin-phase with the effect of the latitudinal differential rotation. For our model, the effect of the radial shear to the net axisymmetric toroidal magnetic field is determined mostly by the near equatorial regions that  are slightly above and below the bottom of the convection zone.
 Also, we find that the toroidal field generation rate depends strongly on the latitudinal profile of the axisymmetric radial magnetic field near the poles. We find that the magnitude of the axisymmetric toroidal flux generation rate in the 3D dynamo model is by about 10 percent higher than in the axisymmetric 2D mean-field dynamo model, due to the bipolar active regions.}
		
	\end{abstract}
	
	
	\section{Introduction}
	
	The total unsigned radial magnetic flux observed in the
		photosphere during the Sun's 11-year activity cycle peaks at about
		$10^{24}$ Mx \citep{Schrijver1984,Schrijver1994}. Some portion of
		this magnetic flux is likely produced by the large-scale
	dynamo in the convection zone. But how and where this flux is produced
	is under debate. A small fraction of it comes from active regions.
	Depending on the complexity of active regions, {the total unsigned radial magnetic field flux associated  with } the bipolar sunspot groups (hereafter, bipolar magnetic regions,
	BMRs) peaks at about $4$-$10\times10^{22}$ Mx \citep{Nagovitsyn2016a,Abramenko2023m}.
	{A part of this flux must participate in the surface flux transport
		and turbulent processes, e.g., via helical convection motion}, contributing
	to the global dynamo. Following the basic ideas of \citet{Parker1955},
	the solar dynamo cyclically transforms the large-scale poloidal
	magnetic field into the toroidal field and back by means of the differential
	rotation and the turbulent $\alpha$ effect caused by the cyclonic
	convective motions.  \citet{CS15} (hereafter
	\citetalias{CS15}) showed that { the hemispheric magnitude of the net axisymmetric toroidal flux production in the bulk of the convection zone  is largely determined by the surface parameters of the differential rotation and the radial magnetic field. Using solar observations, they found that the toroidal flux generation rate by the latitudinal rotational shear is about $10^{23}$~Mx/yr or $10^{15}$~Mx/s. Therefore, this effect is capable of producing the toroidal flux of $10^{24}$ Mx during the solar cycles.} It is interesting to compare this number with the typical unsigned radial magnetic field flux emergence rate of BMRs, which  is about $10^{18}$~Mx/s \citep{Golovko98}.
	Following the arguments of \citetalias{CS15}, it is tempting to conclude
	that the poloidal magnetic flux produced 
	by the magnetic flux of bipolar magnetic regions (BMRs) is sufficient
	for maintaining the solar dynamo. Indeed, this idea has been implemented in the Babcock-Leighton flux transport
	dynamo models \citep{Babcock1961,Leighton1969,Hazra2023a,CS2023S}.  It is worth noting that the  evidences found by \citetalias{CS15} reflect effects of the global axisymmetric dynamo inside the Sun. Our aim is to analyze these effects using the 3D mean-field dynamo model developed by \citet{P22,PKT23} (hereafter, \citetalias{P22,PKT23}). 
	
	Despite the simplicity and visual appeal of the magnetic
	`butterfly' diagram reproduced by the Babcock-Leighton flux transport
	dynamo models, this dynamo scenario is not supported by the
	global-Sun 3D MHD simulations (e.g., \citep{Guerrero2016,Kapyla2016,Warnecke2018,Warnecke2021}.
	The simulations  favor  the dynamo scenario suggested by \citet{Parker1955},
	in which a substantial portion of the poloidal magnetic flux is generated
	by helical turbulence deep in the convection zone. The Parker's dynamo scenario suggests that the solar magnetic activity represents itself in the form
		of the dynamo waves propagating in the convection zone along the isorotation surfaces.

		The helioseismic observations support Parker's idea as well, showing
	the migrating zonal flows (`torsional oscillations'), which reveal
	a dynamical pattern corresponding to the dynamo waves predicted by
	Parker's theory \citep{Kosovichev2019}. The dynamical mean-field
	model developed by \citet{Pipin2018b} and \citet{PK19,Pipin2020}
	reproduced  the observed properties of the torsional oscillations and its` extended' 20-year mode.
		{The  `extended' cycle is observed in the evolution of the ephemeral active regions, the coronal green-line emission, the surface torsional oscillation \citep{Wilson1988}, and the evolution of the low-degree harmonics of the global magnetic fields \citep{ObrShiSok2023}}. The `extended' cycle can be a result of  the extended mode of the axisymmetric  dynamo waves, and the modulation of the heat and angular momentum
		transport in the solar convection zone by the dynamo-generated magnetic fields \citep{Stenflo1994,PK19}. The model also showed  the extended 22-year cycle in the time-latitude variations of the meridional circulation,
	which were found  by helioseismic observations \citep{Komm2018,Getling2021}.
 
	Further developments of our mean-field model included modeling the
	formation and emergence of BMRs driven by magnetic buoyancy instability (\citetalias{P22,PKT23}).
 {Comparing results of the 3D dynamo model runs with the 2D axisymmetric dynamo model, \citetalias{P22} found that the  BMRs' activity may increase of the total toroidal magnetic field flux in the convection zone by about 10 percent. Also, the results of  \citetalias{P22} and \citetalias{PKT23} showed that the surface unsigned radial magnetic field flux reaches a magnitude of $5-7 \times 10^{23}$ Mx in the magnetic cycle maximum. Most of this
flux originates from the emergence and evolution of the BMRs. 
 {\it While the role of the surface BMR activity in dynamo is subdominant, the 3D model  combines the essential elements
	of the Parker and Babcock-Leighton dynamo scenarios.} }
	
{In this paper, we investigate the  toroidal magnetic flux budget for  the {\it axisymmetric} part of the dynamo model using the results of the  3D mean-field data-driven dynamo model of Solar Cycles 23 and 24 (\citetalias{PKT23}).  We compare the modeling results with the corresponding synoptic observations of magnetic fields from the Kitt Peak Observatory (KPO) and the Solar Dynamics Observatory (SDO/HMI)}.
	Section \hyperref[sec:dm]{2} briefly describes the model. Section \hyperref[sec:res]{3} presents the
	results for the toroidal magnetic flux budgets and comparison
	with the synoptic observations. Finally, a summary of the main
	results is presented in Section \hyperref[dis]{4}. Some additional details of the BMR model, the basic dynamo model parameters, and the magnetic field evolution are described in Appendixes \hyperref[app:A]{A} and \hyperref[app:B]{B}.
	
	\section{Dynamo Model}\label{sec:dm}
	
	We employ the dynamo model recently developed 
	by \citetalias{P22}. In the model, we decompose
	the mean magnetic field, $\left\langle \mathbf{B}\right\rangle$, (associated with the  statistical ensemble average) into the sum of the axisymmetric, $\overline{\mathbf{B}}$, and the nonaxisymmetric parts, $\tilde{\mathbf{B}}$, 
	\begin{eqnarray}
		\left\langle \mathbf{B}\right\rangle  & = & \overline{\mathbf{B}}+\tilde{\mathbf{B}}.\label{eq:b0}
	\end{eqnarray}
{These parts are further decomposed into a sum of the poloidal and toroidal components, as follows  \citep{Moss1992}: 
\begin{eqnarray}
\mathbf{\overline{B}} & = & \hat{\mathbf{\phi}}B+\nabla\times\left(A\hat{\mathbf{\phi}}\right)\,,\label{eq:b1}\\
\tilde{\mathbf{B}} & = & \mathbf{\nabla}\times\left(\mathbf{r}T\right)+\mathbf{\nabla}\times\mathbf{\nabla}\times\left(\mathbf{r}S\right),\label{eq:b2}
\end{eqnarray}
where $\hat{\mathbf{\phi}}$ is the azimuthal unit vector, $\mathbf{r}$
is the radius vector, and $\theta$ is
the polar angle. The gauge transformation for potentials $T$ and $S$ involves a sum with arbitrary r-dependent functions (\citealt{Krause1980}). The gauge dependence is removed if the integrals of  $T$ and $S$ over the solid angle are zero.
We seek the solution for ${T}$ and ${S}$ using the spherical harmonic
decomposition. In this case, by definition, the $m=0$ modes in $T$ and $S$
are excluded, and the representation of  Eq.(\ref{eq:b2}) 
is gauge invariant (also, see  \citealp{Berger2018}).} 
 
	The evolution of the large-scale magnetic field is governed by the mean-field induction equation,
	\begin{eqnarray}
	\partial_{t}\left\langle \mathbf{B}\right\rangle  & = & \mathbf{\nabla}\times\left(\mathbf{\boldsymbol{\mathbf{\mathcal{E}}}}+\mathbf{\boldsymbol{\mathbf{\mathcal{E}}}^{\mathrm{(BMR)}}}+\left\langle{\mathbf{U}}\right\rangle\times\left\langle \mathbf{B}\right\rangle \right), \label{eq:dyn}
	\end{eqnarray}
{where   $\boldsymbol{\mathbf{\mathcal{E}}}=\left\langle \mathbf{u\times b}\right\rangle $
	is the mean electromotive force; $\mathbf{u}$ and $\mathbf{b}$ are  randomly fluctuating velocity and magnetic field components. Their evolution equations are solved separately, either analytically or numerically using the so-called test-field method \citep{Schrinner2011}. Typically, such a solution includes nonlinear effects that stem from effects of the global rotation and fluctuations of $\left\langle \mathbf{B}\right\rangle$ caused by background turbulent motions and small-scale dynamo \citep{Maarit2022,BRetal23}.}  We employ the phenomenological term $\mathbf{\boldsymbol{\mathbf{\mathcal{E}}}^{\mathrm{(BMR)}}}$  to describe  that emergence  of  bipolar magnetic regions on the solar surface. 

{In our model, we assume that the large-scale flow is axisymmetric, $\left\langle \mathbf{U}\right\rangle \equiv\overline{\mathbf{U}}$.
The same assumption is applied to the mean entropy and the other thermodynamic parameters.  However, the  magnetic field is three-dimensional. To get the dynamo equations for the axisymmetric toroidal magnetic field evolution, we take the scalar product of Eq.(\ref{eq:dyn}) with the unit vector $\hat{\phi}$. We do the same for the uncurled version  of Eq.(\ref{eq:dyn}) to get the equation for the vector potential, $A$. To obtain the evolution equations for the nonaxisymmetric magnetic field, we  apply the curl and double curl operations to  Eq.(\ref{eq:dyn}). Then, we take the scalar product of these equations with vector $\mathbf{r}$ (see details in \citealt{Krause1980}, and \citealt{Moss1992}).}

  We define $\boldsymbol{\mathcal{E}}$
	using the mean-field MHD magnetohydrodynamics framework of \citet{Krause1980};
	it reads
	\begin{equation}
		\mathcal{E}_{i}=\left(\alpha_{ij}+\gamma_{ij}\right)\left\langle B\right\rangle _{j}-\eta_{ijk}\nabla_{j}\left\langle B\right\rangle _{k},\label{eq:emf}
	\end{equation}
	where $\alpha_{ij}$ describes the turbulent generation by kinetic
	and magnetic helicity, $\gamma_{ij}$ describes the turbulent pumping, and
	$\eta_{ijk}$ is the eddy magnetic diffusivity tensor.  
The $\alpha$ effect tensor includes effect of the magnetic helicity conservation (\citealp{Kleeorin1982,Kleeorin1999}),
\begin{eqnarray}
\alpha_{ij} & = & C_{\alpha}\psi_{\alpha}(\beta)\alpha_{ij}^{\rm K}
+\alpha_{ij}^{\rm M}\psi_{\alpha}(\beta)\frac{\left\langle \mathbf{a}\cdot\mathbf{b}\right\rangle \tau_{c}}{4\pi\overline{\rho}\ell_{c}^2}.\label{alp2d}
\end{eqnarray}
Here, the pseudoscalar $\left\langle \mathbf{a}\cdot\mathbf{b}\right\rangle$ (where $\mathbf{b}=\nabla\times\mathbf{a}$) is 
the small-scale magnetic helicity density, $C_{\alpha}$ is the dynamo parameter characterizing the magnitude
of the hydrodynamic $\alpha$ effect, and $\alpha_{ij}^{\rm K}$ and
$\alpha_{ij}^{\rm M}$ are the anisotropic versions of the kinetic and magnetic $\alpha$ effects (\citealp{Pipin2008a,PK19,BRetal23}).
The radial profiles of $\alpha_{ij}^{H}$ and
$\alpha_{ij}^{M}$ depend on the mean density stratification and  the spatial profiles
of the convective velocity $u_c$, and on the Coriolis number,
\begin{equation}
  {\rm Co} = 2\Omega_0 \tau_c, \label{eq_M8}
\end{equation}
where $\Omega_{0}$ is the global angular velocity of the star and
$\tau_{c}$ is the convective turnover time. 
The magnetic quenching function $\psi_{\alpha}(\beta)$ depends on
the parameter $\beta=|\langle\mathbf{B}\rangle|/\sqrt{4\pi\overline{\rho} u^2_c}$ \citep{PK19}.  The evolution of magnetic helicity 	is governed by the equation of the integral balance of the helicity density,
	\begin{equation}
	\left(\frac{\partial}{\partial t}+\boldsymbol{\overline{\mathbf{U}}\cdot\nabla}\right)\left\langle \chi\right\rangle ^{(\mathrm{tot})}=-\frac{\left\langle \mathbf{a}\cdot\mathbf{b}\right\rangle }{R_{m}\tau_{c}}+2\eta_{0}\left\langle \mathbf{B}\right\rangle \cdot\left\langle \mathbf{J}\right\rangle +\mathbf{\nabla\cdot}\eta_{\chi}\mathbf{\nabla}\left\langle \mathbf{a}\cdot\mathbf{b}\right\rangle, \label{helab}
	\end{equation}
	where $\left\langle \chi\right\rangle ^{(tot)}=\left\langle \mathbf{a}\cdot\mathbf{b}\right\rangle +\left\langle \mathbf{A}\right\rangle \cdot\left\langle \mathbf{B}\right\rangle $
	is the total magnetic helicity, { $\eta_{0}$ is the constant microscopic diffusivity, $\left\langle \mathbf{J}\right\rangle=\nabla\times \left\langle \mathbf{B}\right\rangle$ is the mean current density, and parameter $R_{m}=u_c\ell_c/\eta_0=10^{6}$
is the magnetic Reynolds number, where $u_c$ and $\ell_c$ are  the RMS convective velocity and the mixing length of the turbulent convection, respectively.  We choose  $\eta_0$  to be 
 small enough so that its amplitude does not affect the evolution of  the small-scale helicity density.} Following the results of \citet{Mitra2010}, we use the mixing-length estimation of the magnetic diffusivity $\eta_{\chi}=\frac{1}{10}\eta_{T}$, { where, $\eta_T=u_c\ell_c/3$. Similarly to $\alpha_{ij}$, the structure and radial profiles  of $\gamma_{ij}$  and $\eta_{ijk}$ are determined by the effects of the global rotation and large-scale magnetic field on turbulent flows (\citealp{Pipin2008a,PK19} and \citetalias{P22}). Figure \ref{fig1} in Appendix \hyperref[app:B]{B} shows the convection zone profiles for $\alpha_{ij}^K$, $\eta_{ijk}$ and the effective drift velocity, which is defined as a sum of the turbulent pumping and meridional circulation  \citep{Warnecke2018}}.  

 {The calculation of the small-scale helicity density evolution from Eq.(\ref{helab}) depends on the gauge of the large-scale magnetic field vector potential, $\left\langle\mathbf{A}\right\rangle$.  The axisymmetric part of the vector-potential satisfies the Coulomb and Poincare gauges, by default. The nonaxisymmetric part of $\left\langle\mathbf{A}\right\rangle$ depends on gauges of scalars $S$  and $T$. In our spherical harmonic decomposition, we exclude the $m=0$ modes of  $T$ and $S$. In this case,  the representation of Eq.(\ref{eq:b2}) 
is gauge invariant (\citealp{Krause1980,Berger2018}).}  
	
	The electromotive force induced by the bipolar magnetic regions is determined by 
	\begin{equation}
		\mathcal{E}_{i}^{(\mathrm{BMR})}=\alpha_{\beta}\delta_{i\phi}\left\langle B\right\rangle _{\phi}+V_{\beta}\left(\hat{\boldsymbol{r}}\times\left\langle \mathbf{B}\right\rangle \right)_{i}.\label{eq:ebmr}
	\end{equation}
{Parameter $\alpha_{\beta}$ determines the magnetic field generation effect, which is related to the BMRs tilt \citepalias{P22}.  The second
	term of this equation models the emergence of the nonaxisymmetric magnetic field  in the form of bipolar magnetic regions. Parameter $V_{\beta}$ is related to the buoyancy velocity of BMRs in the convection zone.} The reader can find further details about $\mathbf{\boldsymbol{\mathbf{\mathcal{E}}}^{\mathrm{(BMR)}}}$  in Appendix and \citetalias{P22,PKT23}. 
	
 The reference profiles of the mean
	thermodynamic parameters, such as entropy, density, temperature, and the parameters of the background turbulent convection, such as the  convective
	turnover time, $\tau_{c}$, and the mixing length, $\ell_{c}$, are
	determined from the solar interior model MESA \citep{Paxton2013}. For this computation, we use the default choice of the MESA input parameters,  which includes the information on the solar mass (1M$_{\odot}$), metalicity (z=0.02), and age (4.6 Gyr). The convective RMS velocity is determined from the mixing-length approximation, 
\begin{equation}
\mathrm{u_{c}=\frac{\ell_{c}}{2}\sqrt{-\frac{g}{2c_{p}}\frac{\partial\overline{s}}{\partial r}},}\label{eq:uc}
\end{equation}
	where $\ell_{c}=\alpha_{\mathrm{MLT}}H_{p}$ is the mixing length,
	$\alpha_{\mathrm{MLT}}=1.9$ is the mixing length parameter, and $H_{p}$
	is the pressure scale height. Equation (\ref{eq:uc}) determines {\it the reference
	profiles} for the eddy heat conductivity, $\chi_{T}$, eddy viscosity,
	$\nu_{T}$, and eddy diffusivity, $\eta_{T}$, as follows, 	\begin{eqnarray}
			\chi_{T} & = & \frac{\ell^{2}}{6}\sqrt{-\frac{g}{2c_{p}}\frac{\partial\overline{s}}{\partial r}},\label{eq:ch}\\
			\nu_{T} & = & \mathrm{Pr}_{T}\chi_{T},\label{eq:nu}\\
			\eta_{T} & = & \mathrm{Pm_{T}\nu_{T}},\label{eq:et}
		\end{eqnarray}
where $\mathrm{Pr}_{T}=3/4$ is the turbulent Prandtl number. We put the magnetic turbulent Prandtl number $\mathrm{Pm}_{T}=10$. This choice allows us to match the solar cycle period in our dynamo model. 

The model includes the effects of the dynamo-generated magnetic field on the evolution of global axisymmetric flows and the heat transport in the solar convection zone. We take into account the effect of the nonaxisymmetric magnetic field using the longitudinal averaging of
the Lorentz force and the Maxwell stresses. The reader can find further details in our papers \citep{PK19,P22,PKT23}.  The large-scale flow and turbulence parameters profiles of our 3D model as well as the modeled surface evolution of the radial magnetic field are presented in Appendix \hyperref[app:B]{B}.

	\subsection{The boundary conditions and parameters}
	
	Our models include the convection zone and the convective overshoot region. The bottom of the
	integration domain is fixed at the bottom of the overshoot zone, $r_{i}=0.67$R, {where $R\equiv R_{\odot}$ is the radius of the Sun,}. The convection zone
	extends from $r_{b}=0.728R$ to $r_{t}=0.99R$.  The top layers of the solar convection zone are not included because of the strong density variations in the very near surface layer, which can be problematic because of the numerical resolution issues. At the bottom boundary, $r_{i}=0.67$R, we set the magnetic field induction vector to zero. At the top boundary, we use the condition,
	\begin{eqnarray}
		\delta\frac{\eta_{T}}{r_{\mathrm{t}}} \overline{B}_{\phi}\left(1+\left(\frac{\left|\overline{B}_{\phi}\right|}{B_{\mathrm{esq}}}\right)\right)+\left(1-\delta\right)\mathcal{E}_{\theta} & = & 0,\label{eq:tor-vac}
	\end{eqnarray}
which allows penetration of the toroidal magnetic field to the surface \citep{Pipin2011a}. Parameter $\delta$ controls the magnitude of the axisymmetric toroidal field at the top boundary. For the set of parameters: $\delta=0.999$
	and $B_{\mathrm{esq}}=5$G,  the surface toroidal field magnitude is
	around 1.5 G. The poloidal magnetic field is potential outside the
	dynamo domain.{ For the nonaxisymmetric  magnetic field, we put  potentials $T$  and $S$ 
 to zero at  the bottom boundary, $r_i$. At the top,  we use the standard vacuum boundary conditions, i.e,  $T=0$,  and  $S$  is defined by  the potential magnetic field solution outside of the dynamo domain.}  The dynamo model equations are solved using the spherical harmonics with angular degree $\ell\le 72$.  For the numerical solution, we employ
	the \textsc{fortran} version of the \textsc{shtns} library of \citet{shtns}. To integrate over the radius, we employ the finite differences  using the uniform mesh of 80 points over $r_{b}-r_t$ interval and 30 mesh points for the region below the convection zone,  $r_{i}-r_b$. The model employs the second-order Crank-Nicholson scheme with alternate directions for integration in time. 
	
	{The full dynamo cycle of about $20$ years is reproduced if the turbulent magnetic Prandtl  number $\mathrm{Pm}_{T}=10$, and  the $\alpha$-effect parameter $C_{\alpha}=0.045$. The critical value of this parameter $C^{cr}_{\alpha}\approx 0.04$. The magnitude of the alpha effect in our model is about 0.5 m/s.   These parameters are adopted in the reference axisymmetric model T0. In the range $C_{\alpha}=0.04 -0.045$, the dynamo solution shows the antisymmetric magnetic field about the equator.  
 The magnetic flux loss due to the mean-field magnetic buoyancy results in  a weakly nonlinear dynamo regime with the maximum toroidal magnetic 	field strength of about 2.5 kG and parameter $\mathrm{\beta=\left|\left\langle \mathbf{B}\right\rangle \right|/\sqrt{4\pi\overline{\rho}u_{c}^{2}}}\le0.2$ 	in the solar convection zone. 
 We evolve the axisymmetric model to a quasi-stationary stage and choose the profile of the magnetic field at the minimum of a magnetic cycle with the sign of the polar magnetic field, which corresponds to the minimum state of Solar Cycle 22 on Jan 1, 1996. This magnetic field profile is considered as the initial state of the data-driven non-axisymmetric model S2. In this model,  a part of the axisymmetric poloidal magnetic field flux is generated due to the evolution of the surface BMRs. Therefore, in this model, we reduced  the baseline  magnitude of the mean-field $\alpha$  effect to the threshold value of the axisymmetric dynamo model ($C_\alpha=C^{cr}_{\alpha}$). To reproduce the magnitude and duration of Solar Cycles 23 and 24, we further decreased $C_\alpha=0.036$ after five years from the start of Cycle 23, and  increased it back to $C_\alpha=C^{cr}_{\alpha}$  at the end of the cycle in 2007, see the further details of the data-driven model S2 in \citetalias{PKT23}. }
	
	To evaluate the effects of BMRs in the magnetic
	flux budget, we compare the non-axisymmetric data-driven model S2 with the axisymmetric
	reference model T0 calculated without BMRs. 
	
Some output parameters of our models are given in Table \ref{tab}, where we use the
	same model notations as  in \citetalias{PKT23}. The model results were extensively discussed in our previous papers \citep{PK19,P22,PKT23}. For convenience, we included some of those results in the Appendices.
	
	\begin{deluxetable*}{cccccc}\label{tab}
		\tablenum{1}
		\tablecaption{Parameters of the dynamo models  \citepalias{PKT23}.}
		\tablewidth{0pt}
		\tablehead{
			\colhead{Model } & \colhead{Baseline} & \colhead{BMR injection} & \colhead{Toroidal flux and flux rate} & \colhead{Poloidal flux and flux rate} &  \colhead{ Cycle period}  \\
			\colhead{} & \colhead{ $C_{\alpha}$} & \colhead{setup} & \colhead{ [Mx], [Mx/yr]} & \colhead{ [Mx], [Mx/yr]} & \colhead{[yr]}   }
		\decimalcolnumbers
		\startdata
		T0 & 1.1$C^{cr}_{\alpha}$ & no injection & $8\times 10^{23}$, $3\times 10^{23}$ & $1.9\times 10^{22}$, $4\times 10^{21}$  & 10.4  \\
		S2 & $C^{cr}_{\alpha}$ & data-driven injection & $10^{24}$, $4\times 10^{23}$ & $1.9\times 10^{22}$, $6\times 10^{21}$  & 11.2, 11.6 \\
		\enddata
		\tablecomments{Model T0 is the axisymmetric baseline model without BMRs. Model S2 is the data-driven
			model with the BMR emergence. The forth column shows the magnitude
			of the unsigned toroidal flux in the convection zone and its generation rate;
			the fourth column shows the magnitude of the unsigned surface axisymmetric
			radial magnetic field flux; the last column shows the duration of the
			activity cycles (half dynamo periods of the magnetic cycles).}
	\end{deluxetable*}
	
	\section{Magnetic Flux Budget}\label{sec:res}
	
	To estimate variations
	of the axisymmetric toroidal magnetic flux in the dynamo region, we follow the
	approach of \citetalias{CS15} and apply Stokes' theorem to the
	induction equation. The time derivative of the axisymmetric toroidal
	magnetic field flux in the Northern hemisphere of the Sun is calculated
	as follows, 
	\begin{equation}
		\frac{\partial\Phi_{\mathrm{tor}}^{\mathrm{N}}}{\partial t}=\oint_{\delta\Sigma} \left(\overline{\mathbf{U}}\times\mathbf{\overline{B}}+\overline{\boldsymbol{\mathcal{E}}}+\overline{\boldsymbol{\mathcal{E}}}^{(BMR)}
  \right)\cdot\mathrm{d\mathbf{l}},\label{eq:St}
	\end{equation}
	where $\Phi_{\mathrm{tor}}^{\mathrm{N}}=\int_{\Sigma}\overline{B}_{\phi}\mathrm{dS}$,
	$\Sigma$ is the area of the meridional cut through the solar convection
	zone in the hemisphere, $\delta\Sigma$ stands for the contour line
	confining the cut, and $\mathrm{d\mathbf{l}}$ is the line element of $\delta\Sigma$;
	$\overline{\mathbf{U}}$, $\mathbf{\overline{B}}$, and $\overline{\boldsymbol{\mathcal{E}}}$, 
 $\overline{\boldsymbol{\mathcal{E}}}^{(BMR)}$
	are the large-scale axisymmetric flow velocity, magnetic field vector, and the longitudinally averaged components of the mean 
	electromotive force. {The integration over contour $\delta\Sigma$ is performed clockwise, first, along the top boundary at
		$r_{t}=0.99$R from the equator to the North pole, then along the North axis to the bottom boundary at
		$r_{i}=0.67$R, and, finally, along the radius in the equatorial plane, back to the top boundary. }  A similar contour integration can
	be written for the Southern hemisphere flux, $\Phi_{\mathrm{tor}}^{\mathrm{S}}$.
	
	Similarly to \citetalias{CS15}, we estimate the RHS of Equation (\ref{eq:St})
	in the coordinate system  co-rotating with the solar equator with the angular velocity  $\Omega^{(E)}=462$ nHz :
	\begin{eqnarray}
		\frac{\partial\Phi_{\mathrm{tor}}^{\mathrm{N}}}{\partial t} & = & \!\!\int_{0}^{\pi/2}\!\overset{\mathrm{I_{1}}}{\overbrace{\left(\!\overline{U}_{\phi}\!-\!\overline{U}_{0\phi}\!\right)\overline{B}_{r}}}r_t\mathrm{d}\theta\!\label{eq:integ}\\
		& - & \!\!\int_{r_i}^{r_{\mathrm{t}}}\overset{\mathrm{I_{2}}}{\overbrace{\left(\overline{U}_{\phi}\!-\!\overline{U}_{0\phi}\right)^{(E)}\overline{B}_{\theta}^{(E)}}}\mathrm{dr}\!\nonumber \\
		& - & \!\int_{r_i}^{r_{\mathrm{t}}}\overset{\mathrm{I_{3}}}{\overbrace{\left(\!\overline{\mathcal{E}}_{r}^{(N)}\!-\!\overline{\mathcal{E}}_{r}^{(E)}\!\right)}}\mathrm{dr}+\!\int_{0}^{\pi/2}\!\overset{\mathrm{I_{4}}}{\overbrace{\left(\!\overline{\mathcal{E}}_{\theta}^{(\mathrm{t)}}r_t\!-\!\overline{\mathcal{E}}_{\theta}^{(\mathrm{i})}r_i\!\right)}}\mathrm{d}\theta.\nonumber 
	\end{eqnarray}
{Here,   $\overline{U}_{0\phi}=R\sin\theta\Omega^{(E)}$ is the rotational velocity, $R\equiv R_{\odot}$, 
	superscripts $^{(E)}$ and $^{(N)}$ refer to the solar equator and North pole; superscripts $^{(t)}$ and $^{(i)}$ denote 	the top boundary level at $r_{t}=0.99$R and the inner boundary level at 	$r_{i}=0.67$R.  The integral kernels, $I_{1}$ and $I_{2}$, 	represent magnetic induction $\overline{\mathbf{U}}\times\mathbf{\overline{B}}$
		calculated along the top boundary and the equatorial radius,  respectively. }
		The kernels, $I_{3}$ and $I_{4}$, represent the contributions of the turbulent electromotive force.  
  
	{The nonaxisymmetric components of the large-scale magnetic field and magnetic field of  BMRs can contribute to this budget equation via the longitudinally averaged effects of the mean-electromotive force and effects of  $\overline{\boldsymbol{\mathcal{E}}}^{(BMR)}$.  In the solar case, our model shows that the axisymmetric toroidal field is dominant inside the convection zone (\citetalias{P22, PKT23}). Therefore, we neglect the direct effects of the nonaxisymmetric magnetic field and BMR's activity in  Eq. (\ref{eq:integ}). Injections of  BMRs in the convection zone and the near-surface activity of the nonaxisymmetric  magnetic field  generate the poloidal magnetic field. These effects are included in the terms   $I_{1}$ and $I_{2}$.}
It is noteworthy that the first two terms of Eq.(\ref{eq:integ})could be transformed using integration by part as follows,
 \begin{eqnarray}
		 & & \int_{0}^{\pi/2}\left(\!\overline{U}_{\phi}-\overline{U}_{0\phi}\!\right)\overline{B}_{r}r_t\mathrm{d}\theta - \int_{r_i}^{r_{\mathrm{t}}}\left(\overline{U}_{\phi}\!-\!\overline{U}_{0\phi}\right)^{(E)}\overline{B}_{\theta}^{(E)} \mathrm{d}r\label{I1I2} \\
	&=& -\int_{0}^{\pi/2} r_{t} A(r_t,\theta\sin\theta\frac{\partial\Omega}{\partial \theta} \mathrm{d}\theta  -\int_{r_i}^{r_t} r A(r,\frac{\pi}{2})\frac{\partial\Omega}{\partial r} \mathrm{d}r\label{I1I2e},
	\end{eqnarray}
where $A$ is the axisymmetric vector-potential, see the Eq(\ref{eq:b1}. While the first line (Eq.{\ref{I1I2}}) is very useful for the numerical estimation, especially using observational data sets, it could be misinterpreted sometimes. The second line preserves the standard interpretation of the the differential rotation effect. Here, the term, $-r_t\sin\theta A(r_t,\theta)$, represents (with the factor 2$\pi$) the net flux of the meridional component of the poloidal magnetic field  across the radial section of the convection zone at polar angle $\theta$. The second term $rA(r,\frac{\pi}{2})$ represents the net flux of the axisymmetric radial magnetic field through the northern hemisphere of radius $r$.

  To estimate $I_{1}$ and $I_{2}$, we use the mean angular velocity profile produced in our dynamo models. In this procedure, we neglect the effect of the torsional oscillations  on $I_{1}$ and $I_{2}$ (associated with the large-scale dynamo). We normalize the kernels by dividing them on the solar radius. 
	
	\begin{figure}
		\centering\includegraphics[width=0.8\textwidth]{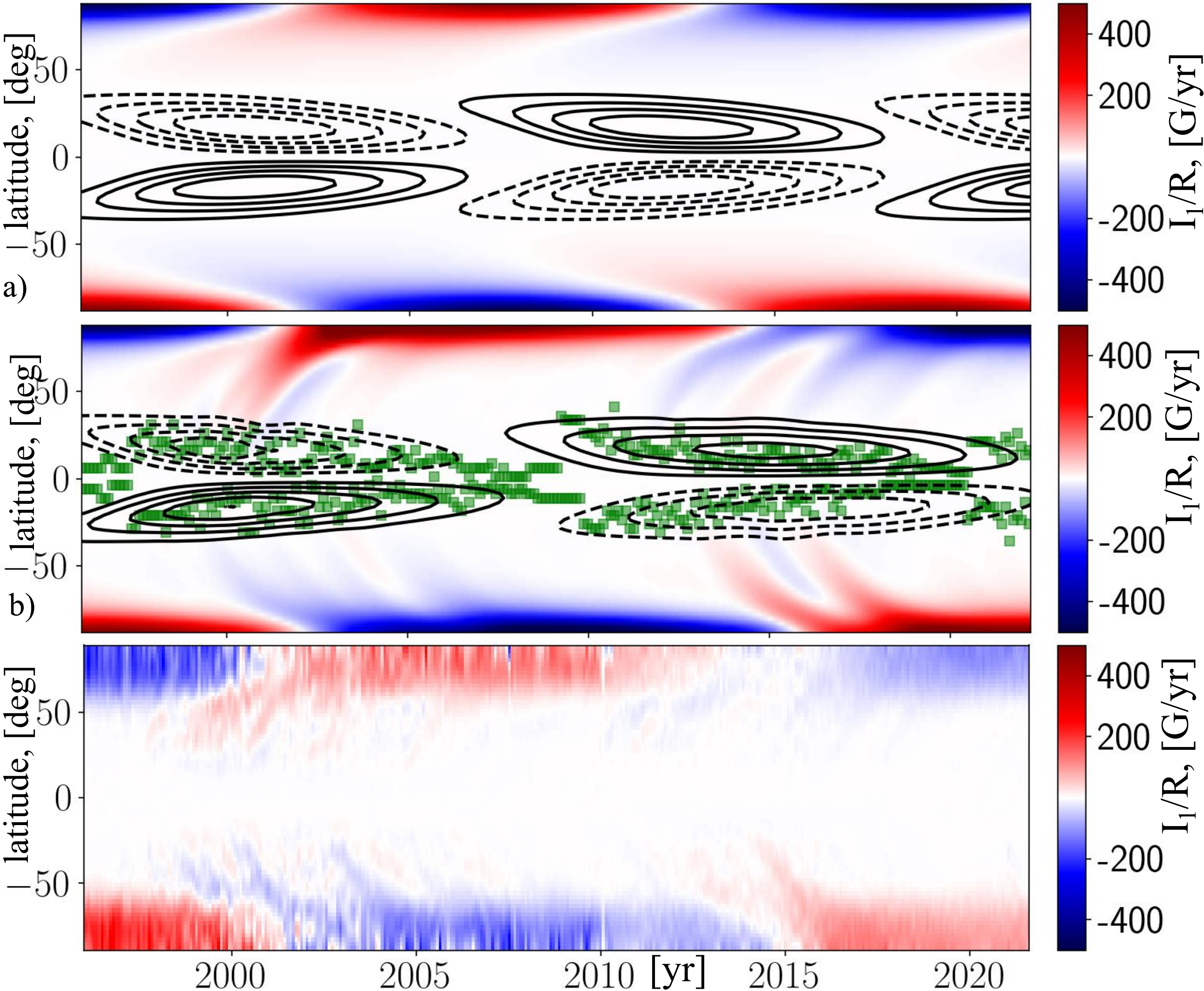}
  \caption{\label{fig:s0} Contour lines in panels (a) and (b) show the time-latitude diagrams of the near-surface toroidal magnetic field in the range of $\pm1$kG for Models T0 and S2; the background color images{ show the generation rate of the toroidal field by } the latitudinal shear , i.e., integral kernel $I_{1}$ of Equation~(\ref{eq:integ}); the green squares in panel (b) mark the latitudinal positions of the BMRs's injections; panel (c) shows the toroidal field generation rate calculated  for the axisymmetric magnetic field using the Kitt Peak Observatory (KPO) and SDO/HMI synoptic maps,{ hereafter $R\equiv R_{\odot}$}.}
	\end{figure}
	
	Figure \ref{fig:s0}(a-c) shows the time-latitude diagrams of the
	integral kernel, $I_{1}$, calculated for our dynamo models and for
	the observation data. 
{To estimate $I_1$ from observational data, we use the same  surface differential rotation profile as in the model.}
 {  We use synoptic maps of the radial magnetic field obtained from the data set produced cooperatively by the NASA Goddard Space Flight Center and NOAO Space Environment Laboratory. This work utilizes the SOLIS data obtained by the National Solar Observatory (NSO) Integrated Synoptic
Program (NISP). This data set is combined with the synoptic maps of the radial magnetic field from SDO/HMI \citep{Sun2011}. }
 
The dynamo models qualitatively agree with the
	observational data, as well as the results of \citetalias{CS15}, regarding the magnitude and sign of $I_{1}$.
 This agreement comes from the similarity of the   axisymmetric
	magnetic field evolution in the dynamo models and observations. {However, in our model, parameter $I_{1}$  is confined to a much narrower region closer to the poles than in the observations. This is because of  a stronger concentration of the radial magnetic field to the poles in our model in comparison to the solar observations.}
	\begin{figure}
		\centering \includegraphics[width=0.85\columnwidth]{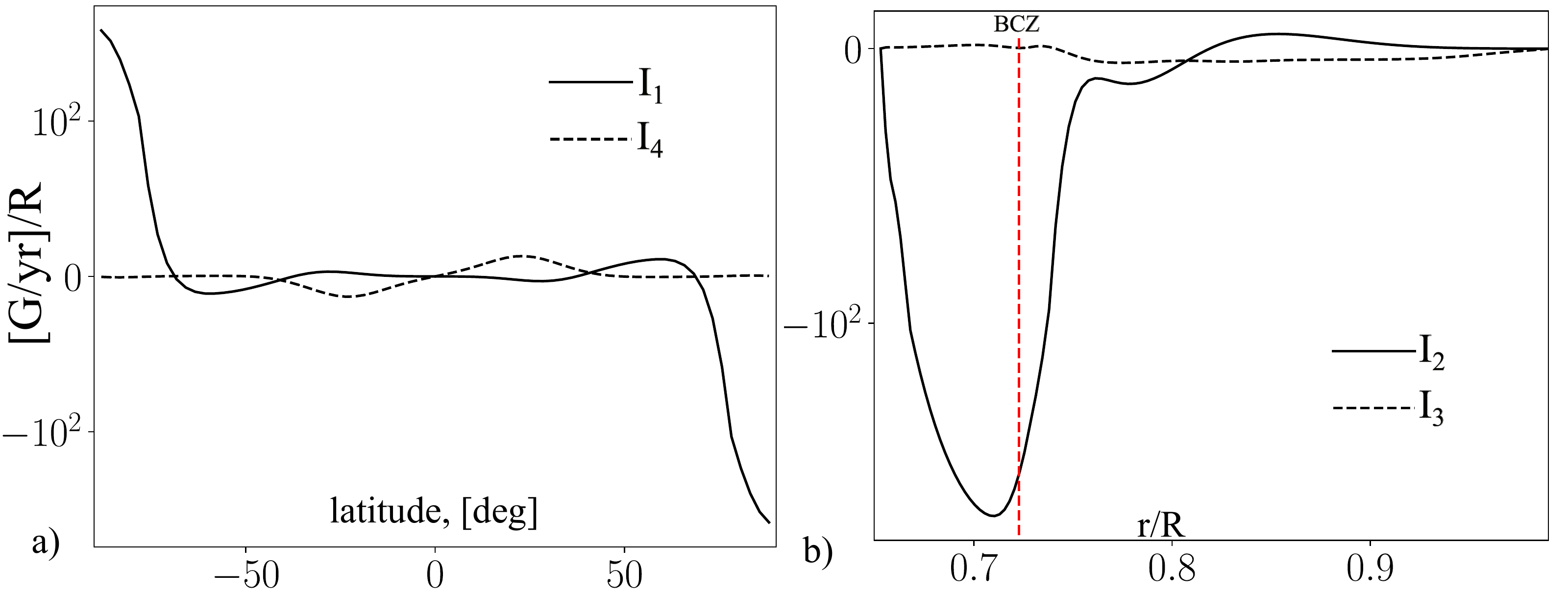} \caption{\label{fig:bdgs}Estimation of contributions to the magnetic flux budget (Equation ~(\ref{eq:integ}))
			for the cycle minimum in 1996. The linear threshold for the y-axis is at 100 G/yr/R. }
	\end{figure}
	
	\begin{figure}
		\centering \includegraphics[width=0.45\textwidth]{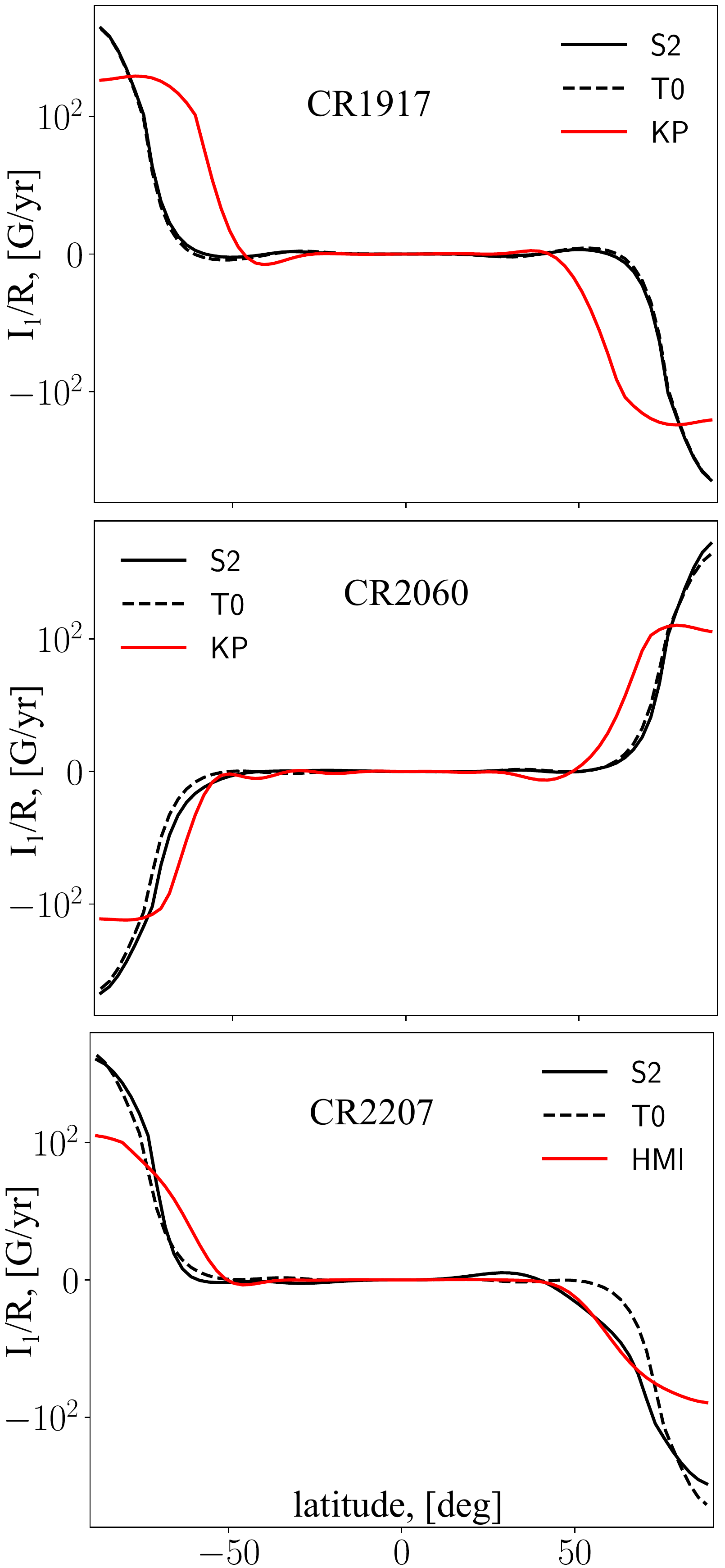}
\caption{\label{fig:kp} 
  Panels (a), (b), and (c) show the latitudinal dependence of the surface term,  $I_{1}$,  to the toroidal
			magnetic flux budget 
			Equation~(\ref{eq:integ}) {for the instances of the activity minima of Cycles 22, 23, and 24} in
			our dynamo models and the observations. {The linear threshold for the y-axis is set at 100 G/yr/R.} } 
	\end{figure}
	
	Figure \ref{fig:bdgs} shows the latitudinal profiles of $I_{1}$ and $I_{4}$ 
	and the radial profiles of $I_{2}$ and $I_{3}$ for model T0 for {the magnetic cycle minimum in 1996}. The models show a sharp poleward increase
	of $I_{1}$ (Figure~\ref{fig:bdgs}(a)). 
 {This effect contributes to the axisymmetric toroidal magnetic field generation by the latitudinal shear in the bulk of the convection zone.}

 {The term $I_{2}$ describes the axisymmetric  toroidal magnetic field generation by the radial gradient of the angular velocity. This term has a maximum near the bottom of the convection
	zone, where it has same sign and nearly the same order of magnitude as $I_{1}$ at the northern pole.} In the
	main part of the convection zone, it changes sign. {The terms $I_3$ and $I_4$ include the effects of the diffusive loss and the loss of the magnetic flux due to magnetic buoyancy. Generally, the sign of these terms is opposite to the sign of the flux generation terms $I_1$ and $I_2$}.
 
 Figure \ref{fig:kp} illustrates the latitudinal profiles
	of $I_{1}$ for the minima of Cycles 22, 23, and 24 in our dynamo
	models and the observations. In all cases, the observations
	show a step-like increase of $I_{1}$ above $50^{\circ}$ latitude
	and almost uniform $I_{1}$ near the solar poles. The dynamo models
	show similar profiles, but the distribution of $I_{1}$
	is not uniform near the poles. \citetalias{PKT23} found that the polar magnetic 	field in solar observations does not increase toward the poles as 	strongly as our models show. This can be a reason for nearly uniform $I_{1}$ near the solar poles for the data of observations.

	\begin{figure}
		\centering \includegraphics[width=0.99\textwidth]{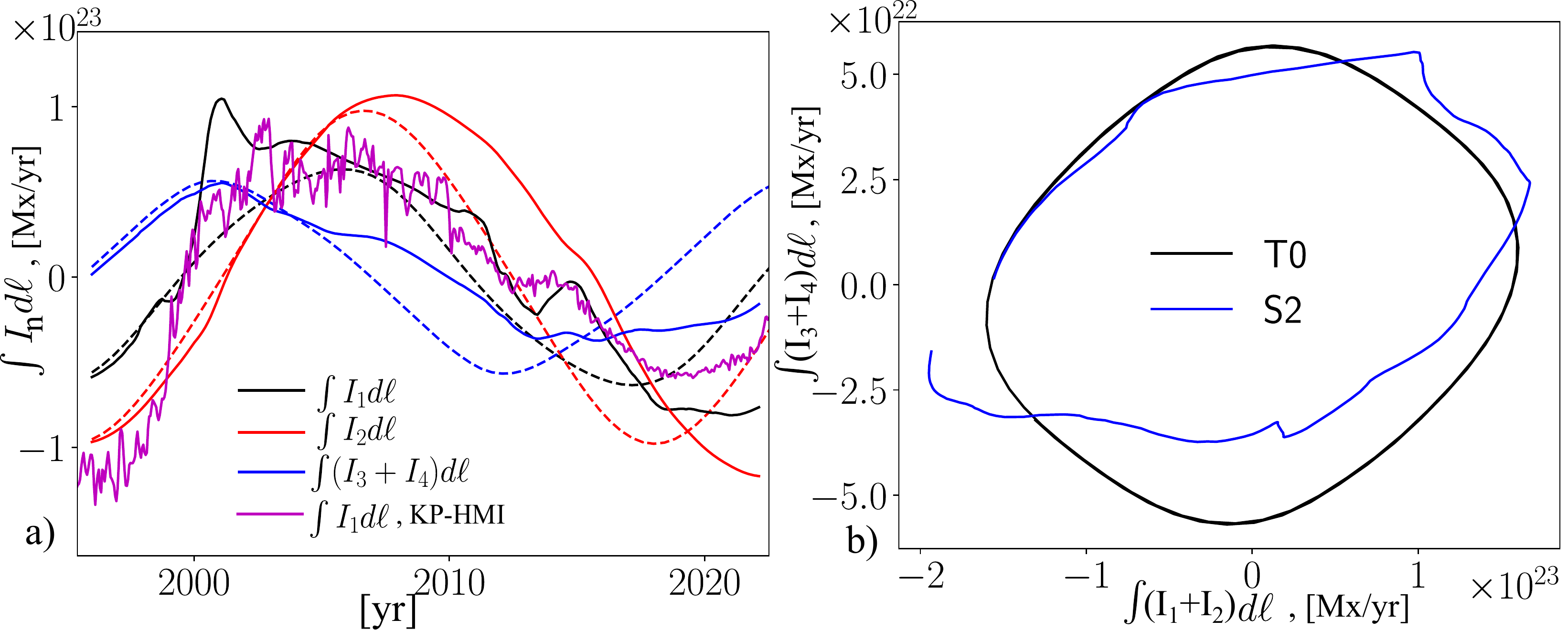}
  \caption{\label{fig:bdg}(a) Time evolution of the integral RHS contributions of Equation (\ref{eq:integ}) for the northern hemisphere
			for models S2 (solid lines) and T0 (dashed lines) and for the observational data set; (b) the phase diagram of the toroidal flux generation  and loss rates
			(the horizontal axis shows the sum of $\int_{0}^{\pi/2} I_1 r_t\mathrm{d}\theta+\int_{r_i}^{r_t}I_2\mathrm{d}r$) and the vertical axis shows the sum of $\int_{r_i}^{r_t}I_3\mathrm{d}r+\int_{0}^{\pi/2} I_4 \mathrm{d}\theta$) for model T0 (black curve)  and model S2 (blue curve).}
	\end{figure}
	
	Figure \ref{fig:bdg}a shows the time evolution of the RHS contributions
	in Equation~(\ref{eq:integ}) for models T0 and S2, and for the observational data set. Notably, these variations in
	the Southern hemisphere have the opposite sign.   Model S2 shows 
 a larger magnitude of of $I_{1,2}$ than model T0.  The magnitude of this increase is about 10 percents of the whole $I_{1,2}$. The larger  $I_{1,2}$ is due to  the effects of BMRs activity on the magnitude of the axisymmetric poloidal magnetic field. Similar effect was found by \citetalias{P22}. The models fail to reproduce the relatively low $\int_N I_1 d\ell$ of the solar cycle 24. Both models show that the integrals of  $I_{1}$ and $I_{2}$ are about the same magnitude, and they evolve sin-phase. These results suggest an important role of the radial rotational shear in the axisymmetric mean-field dynamo. 
 
 Figure \ref{fig:bdg}b
	shows the phase diagrams illustrating the toroidal flux
	generation rate (horizontal axis) and the loss rate (vertical axis)
	in our dynamo models. The model parameters are similar to those 
	deduced by \citetalias{CS15} from solar observations. Some differences
	are because of the additional generation and loss terms included in
	our models. Also, the behavior of the radial magnetic field near the
	poles affects the magnetic flux budget considerably.
 	
	\begin{figure*}
		\centering \includegraphics[width=0.85\textwidth]{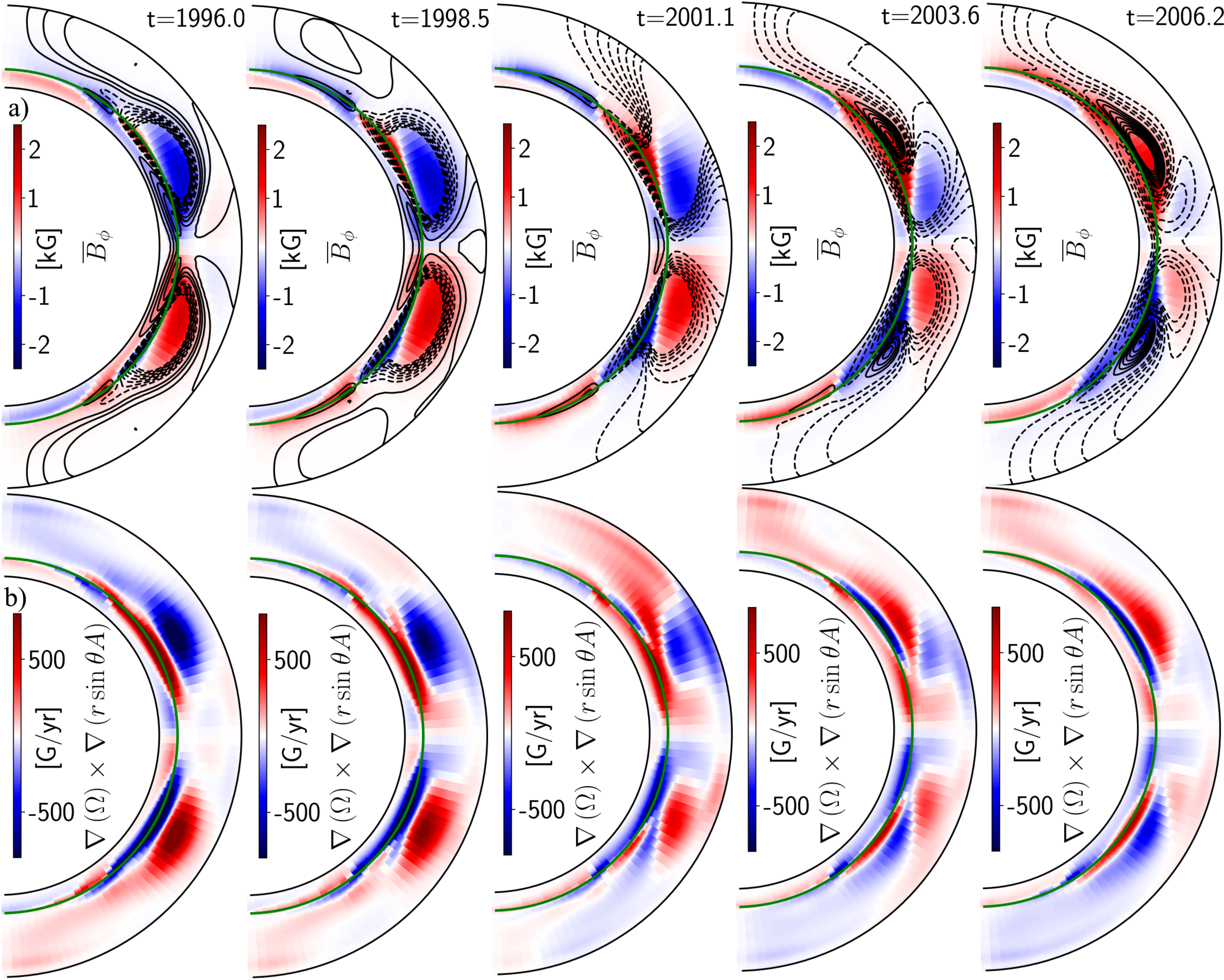}\caption{\label{fig:rlsh}(a) The evolution of the axisymmetric toroidal magnetic
			field (color images) and the poloidal magnetic field{ streamlines (contours, solid lines show the clockwise directed magnetic field)
			during Cycle 23 in model S2, illustrating the dynamo wave propagation
			in the convection zone; the boundaries of the dynamo domain are shown by solid black lines, the green line marks the  bottom of the convection zone}; (b) {the 
				toroidal field generation rate by the differential rotation, i.e., $\boldsymbol{\nabla}\left(\Omega\right)\times\boldsymbol{\nabla}\left(r\sin\theta A\right)$
			(The online version shows the animation of this Figure. The left and right columns of the animation illustrate variations of the axisymmetric magnetic field parameters (panels a) and the 
				toroidal field generation rate (panels b) for Cycles 23 and 24 in Model S2 )}.}
	\end{figure*}
	
	{Figure \ref{fig:rlsh} shows the axisymmetric toroidal magnetic
		field (images in the top panels),  the poloidal magnetic field lines (the solid lines show the clockwise oriented field), and the corresponding toroidal
		field generation rate by the differential rotation, i.e., $\boldsymbol{\nabla}\left(\Omega\right)\times\boldsymbol{\nabla}\left(r\sin\theta A\right)$ (bottom panels) in the convection zone in model S2 during Cycle 23. The toroidal
		field generation rate reaches its maximum of about 2 kG/yr at mid-latitudes near the bottom of the convection zone.
	These images and the accompanying animation illustrate the dynamo
	wave propagation and  the induction  effects of the radial and latitudinal gradients of the angular velocity{ on the evolving large-scale magnetic field.}
 In the lower part of the convection zone, the radial gradient of the angular
	velocity stretches the closed parts of the axisymmetric poloidal magnetic
	field lines. This results in the opposite signs of the toroidal flux generation
	rate in the lower and upper parts of the convection zone. Simultaneously,
	it produces the dynamo wave propagating along the lines of constant rotation according to the Parker-Yoshimura rule \citep{Parker1955,Yoshimura1975}. The latitudinal
	shear stretches the open poloidal field lines in the upper half
	of the dynamo domain. The positive sign of the $\alpha$-effect in
	the Northern hemisphere results in a coordinated action of the radial
	and latitudinal shear in the upper part of the convection zone. It
	follows that the radial gradient of the solar differential rotation provides a substantial part of the
	toroidal flux inside the convection zone and affects the magnitude
	of the mean toroidal flux there. It is noteworthy that, similar to results presented in Fig.\ref{fig:bdgs} and Fig.\ref{fig:bdg}a effects of the differential rotation near the surface of the solar poles and near equator below the bottom of the convection zone have the same sign.
 The open poloidal flux and
	the latitudinal rotational shear control the magnitude of the subsequent
	cycle. Their properties affect the correlation of the polar magnetic field strength during the solar minima
	with the magnitude of the toroidal field dynamo wave of the subsequent
	magnetic cycle.

	\section{Discussion and Conclusions}\label{dis}
	
	The solar dynamo is responsible for the generation and decay of the axisymmetric toroidal and poloidal magnetic field fluxes on the Sun. Motivated by the results of
	\citet{CS15} (CS15), we analyzed the dynamo magnetic flux budget
	using solar synoptic observations and results of our mean-field dynamo models \citep{PKT23}.
	One of the hypotheses of \citetalias{CS15} is the dominant role of
	the surface latitudinal differential rotation in the generation of the axisymmetric
	toroidal magnetic field of the Sun. Our results suggest that  the radial rotational shear can be important as well, at least for the mean-field type distributed dynamo models. The study  shows the generation rate
of the toroidal field by the differential rotation in the 3D dynamo models, which include the emergence of Bipolar Magnetic Regions (BMRs), is higher than in the corresponding axisymmetric 2D models without BMRs. This is because the BMRs activity provides the additional source of the poloidal magnetic field  generation for this type of dynamo (\citetalias{P22}).
	
	We find that the toroidal flux generation rate by the surface latitudinal differential
	rotation strongly depends on the radial magnetic field profile in
	the polar regions. Observations show that the radial magnetic field
	is almost uniform near the solar poles during epochs of the solar minima. In contrast, the dynamo models show a sharp increase in the toroidal flux
	generation rate toward the poles. One reason could be a strong effect of the meridional circulation on the near-surface radial magnetic field. The toroidal flux generation rate
	by the latitudinal differential rotation reaches its maximum at the poles in
	the observations and dynamo models during the solar activity
	minima. This property explains the relative success of the correlation
	between the polar magnetic field strength and the magnitude of the
	subsequent magnetic cycle for the solar cycle predictions \citep[e.g.,][]{Schatten1978a,Choudhuri2007}.
	Our results suggest that this correlation may depend on the
	profile of the radial magnetic field near solar poles. The long-term measurements of the large-scale solar magnetic fields in the solar polar
	regions can help to resolve the issue.}
	
	{Acknowledgments}
	
VP thanks the financial support of the Ministry of Science and Higher
	Education of the Russian Federation (Subsidy No.075-GZ/C3569/278).
	AK acknowledges the partial support of the NASA grants: NNX14AB70G,
	80NSSC20K0602, 80NSSC20K1320, and 80NSSC22M0162.
	
	Data Availability Statements. The data underlying this article are
	available by request.
	
	\bibliographystyle{aasjournal}
	\bibliography{dyn}

	\section{Appendix\label{App}}
	
	\subsection{A. Data-Driven BMR Model\label{app:A}}
	
	Injection of the bipolar magnetic regions is determined by 
	\begin{equation}
		\mathcal{E}_{i}^{(\mathrm{BMR})}=\alpha_{\beta}\delta_{i\phi}\left\langle B\right\rangle _{\phi}+V_{\beta}\left(\hat{\boldsymbol{r}}\times\left\langle \mathbf{B}\right\rangle \right)_{i},\label{eq:ebmr-1}
	\end{equation}
	where the first term describes the $\alpha$-effect caused by the
	BMR tilt, and the second term models the magnetic buoyancy instability.
	The magnetic buoyancy velocity, $V_{\beta}$, includes the turbulent
	and mean-field buoyancy effects \citep{Kitchatinov1992,Kitchatinov1993,Ruediger1995}:
	\begin{eqnarray}
		V_{\beta} & = & \frac{\alpha_{\mathrm{MLT}}u_{c}}{\gamma}\beta_{m}^{2}\mathcal{H}\left(\beta_{m}\right)\xi_{\beta}(t,\boldsymbol{r})\label{eq:bu}
	\end{eqnarray}
	where,  function $\xi_{\beta}$, defines the location and formation of the
	unstable part of the magnetic field, the function $\mathcal{H}\left(\beta\right)$ describes effects
	of magnetic tension, 
 	\[
	\mathcal{H}\left(\beta\right)=\frac{1}{8\beta^{4}}\left(\frac{3}{\beta}\arctan\left(\beta\right)-\frac{\left(5\beta^{2}+3\right)}{\left(1+\beta^{2}\right)^{2}}\right),
	\]
where, $\mathrm{\beta=\left|\left\langle \mathbf{B}\right\rangle \right|/\sqrt{4\pi\overline{\rho}u_{c}^{2}}}$,
	($\mathcal{H}\left(\beta\right)\sim1$ for $\beta\ll1$) and subscript 	'm' marks unstable points. 
	To determine the radial position of the unstable point, $r_m$, we use the maximum of product $\left|\overline{B}\right|I_{\beta}\left(r,\theta\right)$. Here, we restrict $r$ to the upper part of the convection zone (above $0.85R$), where $I_{\beta}$ 	reads,
	\begin{equation}
		I_{\beta}=-r\frac{\partial}{\partial r}\log\frac{\left|\overline{B}\right|^{\zeta}}{\overline{\rho}},\label{eq:inst}
	\end{equation}
	where $\overline{B}$ is the strength of the axisymmetric toroidal
	magnetic field and $\overline{\rho}$ is the density profile. For
	the case of $\zeta=1$, we get Parker's instability condition \citep{Parker1979}.
	In our model, we use $\zeta=1.2$. In the unstable region, we must have $I_{\beta}>0$. 
	We define  $\xi_{\beta}$ as follows, 
	\begin{eqnarray}
		\!\xi_{\beta}\left(\boldsymbol{r},t\right)\! & = & C_{\beta}\negthinspace\tanh\mathrm{\left(\frac{t}{\tau_{a}}\right)}\exp\left(-m_{\beta}\left(\sin^{2}\!\left(\!\frac{\phi-\phi_{m}}{2}\!\right)\right.\right.\!\label{xib}\\
		&  & \left.\left.+\!\sin^{2}\!\left(\!\frac{\theta-\theta_{\mathrm{m}}}{2}\!\right)\!\right)\right)\psi(r,r_{m}),\,t<\delta\mathrm{t_{0}}\nonumber \\
		& = & 0,\,t>\delta t.\nonumber 
	\end{eqnarray}
	where $\psi$ is a kink-type function of radius, 
	\begin{eqnarray}
		\psi(r,r_{m})\! & =\!\! & \frac{1}{4}\left(\!1\!+\!\mathrm{erf}\left(100\frac{\left(r-r_{m}\right)}{R}\right)\!\right)\label{eq:step}\\
		& \times & \left(\!1\!-\!\mathrm{erf}\left(100\frac{\left(r-(r_{m}+0.1R)\right)}{R}\right)\!\right),
	\end{eqnarray}
The latitudinal and longitudinal coordinates $\theta_{m}$
	and $\phi_{m}$ in function $\xi_{\beta}$ are taken from the active
	region database. We relate the BMR areas to the parameter
	$m_{\beta}$ as follows, 
	\[
	m_{\beta}=\frac{2}{\sqrt{\mathrm{S_{a}}/10^{6}}},
	\]
	where $\mathrm{S_{a}}$ is the maximum observed area (in the millionth
	of the hemisphere) of the bipolar active regions. In our computations,
	we exclude BMRs with $\mathrm{S_{a}}<50$. 	
	The solar observations show a wide range of variations in the BMR
	emergence time, $\delta t$ and growth rate, $\tau_a$. Also, there are periods of simultaneous
	emergence of several BMRs in one hemisphere. To avoid the overlaps,
	we reformat the emergence initiation time as follows. Firstly, for
	such cases, we shift the emergence of subsequent BMRs by two time steps
	after the end of the previous BMR emergence. Secondly, we define the
	minimal emergence time $\delta\mathrm{t_{min}}=2$ days, \citep{Weberetal2023}, and assume
	that these BMRs have growth of $\tau_{0}=1$ day.
	Specifically, we define: 
	\[
	\mathrm{\tau_{a}}=\frac{1}{2}\frac{\mathrm{\delta t_{a}}}{\delta\mathrm{t_{min}}}\tau_{0},
	\]
	where $\mathrm{\delta t_{a}}$ is the total emergence time of the
	BMRs, $\mathrm{\tau_{a}}$ is the growth rate. The parameters $S_{a}$,
	$\theta_{m}$ and $\phi_{m}$ are taken from the NOAA database of
	solar active regions (https://www.swpc.noaa.gov/).

	The $\alpha$-effect of the $\boldsymbol{\mathcal{E}}^{(\mathrm{BMR})}$
	is given as follows 
	\begin{equation}
		\alpha_{\beta}=C_{\alpha\beta}\left(1+\xi_{\alpha}\right)\cos\theta V_{\beta}\psi_{\alpha}(\beta).\label{eq:ab}
	\end{equation}
	Here, we put the amplitude of the $\alpha$-effect to be determined
	by the local magnetic buoyancy velocity. The $\xi_{\alpha}$ parameter
	controls the random fluctuation of the BMR tilt. The parameter,
	$C_{\alpha\beta}$ controls the mean amplitude of the BMR $\alpha$-effect
	and tilt. To reproduce Joe's law, we put $C_{\alpha\beta}=0.5$ \citepalias{PKT23}.
	{Note that, for $C_{\alpha\beta}=0$, the BMRs are oriented
		along the axisymmetric toroidal field. } We model
	the randomness of the tilt using the parameter $\xi_{\alpha}$. Similar
	to \citet{Rempel2005c}, the $\xi_{\alpha}$ evolution follows the equation, 
	\begin{eqnarray}
		\dot{\xi}_{\alpha} & = & -\frac{2}{\tau_{\xi}}\left(\xi_{\alpha}-\xi_{1}\right),\label{xia}\\
		\dot{\xi}_{1} & = & -\frac{2}{\tau_{\xi}}\left(\xi_{1}-\xi_{2}\right),\nonumber \\
		\dot{\xi}_{2} & = & -\frac{2}{\tau_{\xi}}\left(\xi_{2}-g\sqrt{\frac{2\tau_{\xi}}{\tau_{h}}}\right).\nonumber 
	\end{eqnarray}
	Here, $g$ is a Gaussian random number. It is renewed every time step,
	$\tau_{h}$ . The $\tau_{\xi}$ is the relaxation time of $\xi_{\alpha}$.
	. The parameters $\xi_{1,2,3}$ are introduced to get smooth variations
	of $\xi_{\alpha}$ .  {Similarly to the above-cited papers, we choose
	the parameters of the Gaussian process as follows: $\overline{g}=0$,
	$\sigma\left(g\right)=1$. We choose $\tau_{\xi}=2$ months, which is roughly equal to the lifetime of big active regions \citep{Nortoneal2023}. The
	$\xi_{\alpha}$ varies independently in the northern and southern
	hemispheres. }
 
 \subsection{B. Dynamo model parameters}\label{app:B}

\begin{figure}
\centering \includegraphics[width=0.8\columnwidth]{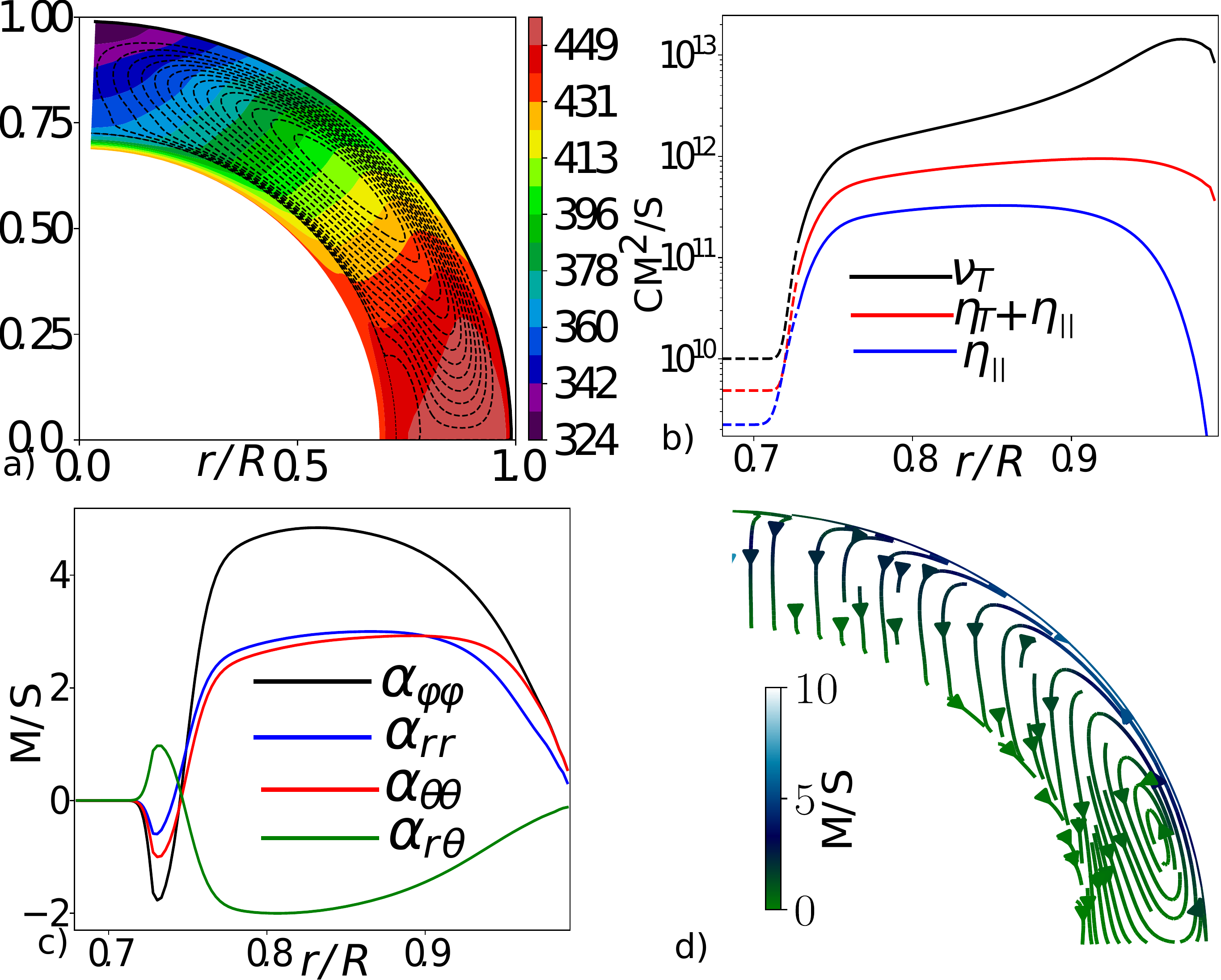} \caption{\label{fig1} a) The meridional circulation (streamlines) and the
angular velocity distributions; the magnitude of circulation velocity
is of 13 m/s on the surface at the latitude of 45$^{\circ}$; b) radial
dependencies of the total, $\eta_{T}+\eta_{||}$, and the rotationally
induced part, $\eta_{||}$, of the eddy magnetic diffusivity and the
eddy viscosity profile, $\nu_{T}$; c) the $\alpha$-effect tensor
distributions at the latitude of 45$^{\circ}$; and d) the streamlines
of the toroidal magnetic field effective drift velocity because of
the meridional circulation and the turbulent pumping effect circulation. Reproduced by
permission from \citetalias{P22}}
\end{figure}

Figure \ref{fig1} illustrates distributions of the angular velocity, meridional circulation, the $\alpha$ - effect, and the eddy diffusivity in the nonmagnetic case model. The amplitude of the meridional circulation on the surface is about 13m/s. The angular velocity distribution is in agreement with the helioseismology data.

\begin{figure}
\centering \includegraphics[width=0.8\columnwidth]{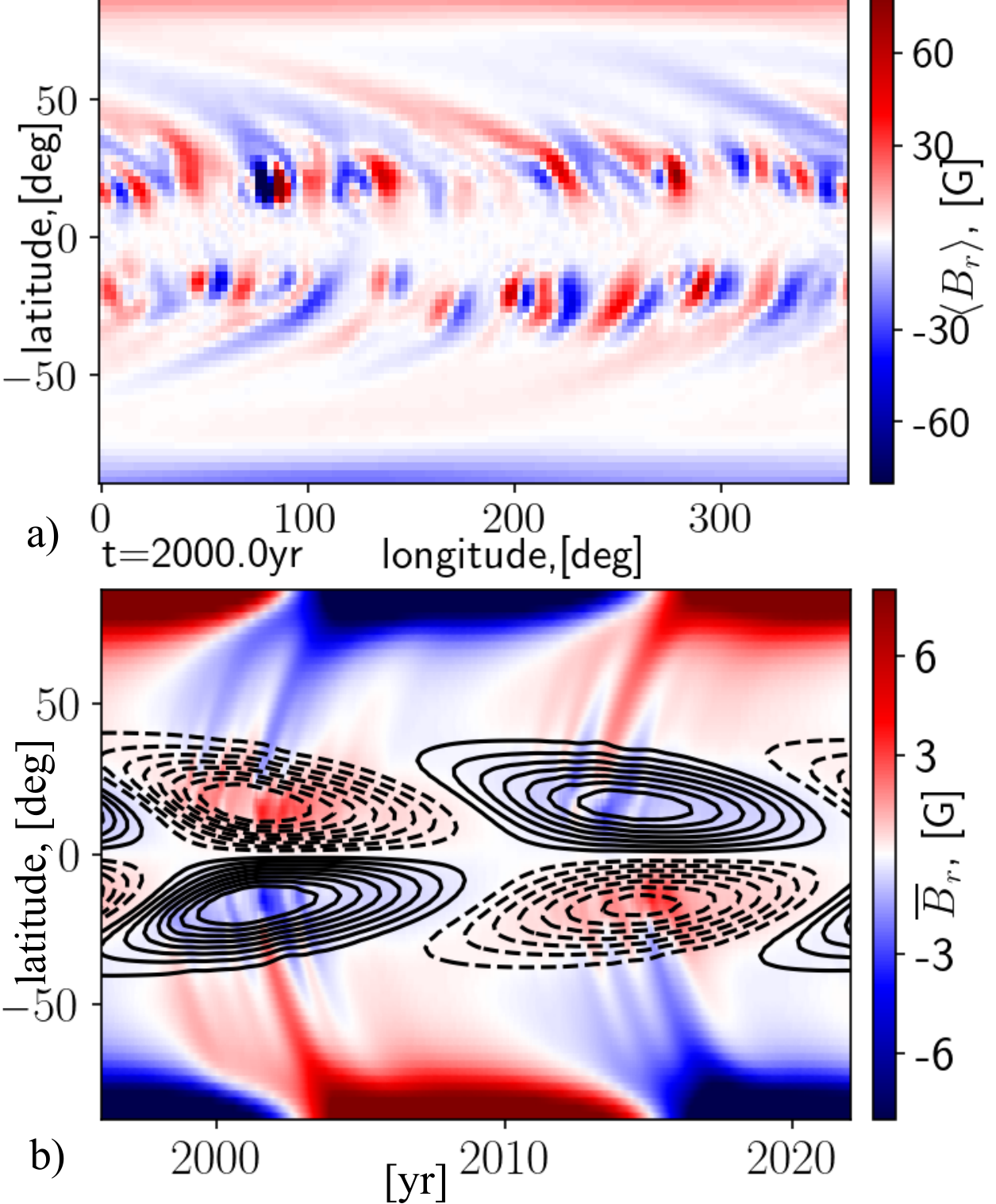} \caption{\label{sn-tl} a) Snapshot of the radial magnetic field,  on the top of the dynamo domainfor the model S2;b)The time-latitude diagram for the axisymmetric toroidal magnetic field (contours in range of $\pm 1kG$) at r=0.9R, and the axisymmetric radial magnetic field at the top is shown by backround color image. The online version shows  the animation of this Figure. Reproduced from 
 the data of the model S2 by permission from \citetalias{PKT23}. }
\end{figure}

Figure \ref{sn-tl}(a) shows a snapshot of the radial magnetic field at the top of the dynamo domain for model S2 for the epoch of the solar maximum around the year 2000.  Figure \ref{sn-tl}(b) shows the time-latitude diagram of the axisymmetric magnetic field in the model S2. Further details can be found in\citetalias{PKT23}.
\end{document}